\documentclass[preprint,12pt]{elsarticle}
\usepackage{graphicx}
\usepackage{booktabs}
\usepackage{amssymb}

\begin{document}
\begin{frontmatter}
\title{Determination of the chemical potential in the Tsallis distribution at LHC energies.}
\author{J. Cleymans$^1$, M.W. Paradza$^{1,2}$}
\address{$^1$ UCT-CERN Research Centre and Department of Physics,\\
University of Cape Town, Rondebosch 7701, Cape, Republic of South Africa\\ 
$^2$ Cape Peninsula University of Technology, Cape Town, South Africa. 
}%
%
\begin{abstract}
The  transverse momentum distributions measured in $p-p$ collisions at the LHC determine the kinetic freeze-out stage 
of the collision. The parameters deduced from these distributions differ from those determined at chemical freeze-out.
The present investigation focuses on the chemical potentials at kinetic freeze-out, these are not necessarily zero 
as they are at chemical freeze-out, the only constraint is that they should be equal for particles and antiparticles at LHC energies.
The thermodynamic variables are determined in the framework of the Tsallis distribution.
The chemical potentials in the Tsallis distribution analysis
of $p-p$ collisions at four different LHC energies have correctly been taken into account.
This leads to a much
more satisfactory analysis of the various parameters and confirms the usefulness of the Tsallis distribution in
high-energy collisions. In particular we find that the temperature $T$ and the volume $V$ at each beam energy are the same 
for all particle types 
considered (pions, kaons and protons).  The chemical potentials for these particles are however very different. Hence we conclude that there is evidence for thermal equilibrium at kinetic freeze-out, albeit in the sense of the Tsallis distribution and
there is no evidence for chemical equilibrium at the final stage of the collision.

\end{abstract}
\end{frontmatter}
%
\section{\label{secIntroduction}Introduction}

The Tsallis distribution~\cite{tsallis} was  first proposed more than 
three decades ago as a generalization of the  Boltzmann-Gibbs distribution and is characterized 
by an additional parameter $q$ which measures the deviation from a standard Boltzmann-Gibbs distribution.
Over the past few  years it has also been applied to high-energy physics where it has been successful in
describing the transverse momentum distributions in $p-p$ 
collisions~\cite{bialas,grigoryan,wong,ristea,biro,parvan1,parvan2,tripathy,zheng1,marques,azmi,sorin,sena,Azmi:2014dwa,Biro:2004qg,Khuntia:2017ite,zheng,de} (for a recent review see e.g.~\cite{Patra:2020gzw}).

One particular form, which satisfies thermodynamic consistency relations~\cite{worku1,worku2} is given by:
\begin{equation} \label{Yield}
  E\frac{d^3N}{d^3p} = gV E\frac{1}{(2 \pi)^3} \left[  1 + ( q - 1) \frac{ E - \mu}{T}\right]^ {-\frac{q}{q -1}},
\end{equation}
where $V$ is the volume, $q$ is the Tsallis parameter and $T$ is the corresponding temperature
or, in terms of variables  commonly used in high-energy physics,   rapidity $y$, transverse mass $ m_T = \sqrt{p_T^2 + m^2}$: 
\begin{equation} \label{YieldNonZeroMu}
\frac{d^2N}{dp_Tdy}  = gV \frac{p_Tm_T}{(2 \pi)^2} \left[  1 + ( q - 1) \frac{m_T \,\cosh\, y - \mu}{T}\right]^ {-\frac{q}{q -1}}.
\end{equation}
In the limit where the parameter $q$ tends to unity one recovers the well-known Boltzmann-Gibbs distribution:
\begin{equation} \label{BG}
\lim_{q\rightarrow 1}\frac{d^2N}{dp_Tdy}  = gV \frac{p_Tm_T}{(2 \pi)^2} \exp\left( - \frac{m_T \,\cosh\, y - \mu}{T}\right).
\end{equation}
The main advantage of Eq.~(\ref{YieldNonZeroMu}) over Eq.~(\ref{BG}) is that it has a polynomial decrease with increasing $p_T$
which is what is obeserved experimentally.

It was recognized early on~\cite{sorin} that there is a redundancy in the number of parameters in this distribution,
namely the four parameters $T,V,q,\mu$ in Eq.~$(\ref{YieldNonZeroMu})$ can be replaced by just three parameters $T_0,V_0,q$ with the 
help of the following transformation:
\begin{eqnarray}
T_0 &=& T \left[1-(q-1) \frac{\mu}{T} \right], \qquad  \mu\leq\frac{T}{q-1},
\label{T0}  \\
V_0 &=& V  \left[1-(q-1) \frac{\mu}{T} \right]^{\frac{q}{1-q}},
\label{V0}
\end{eqnarray}
leading to a 
transverse momentum distribution which can thus be written equivalently as
\begin{equation} \label{YieldIndexZero}
\frac{d^2N}{dp_Tdy}  = gV_0 \frac{p_Tm_T}{(2 \pi)^2} \left[  1 + ( q - 1) \frac{m_T \,\cosh\, y}{T_0}\right]^ {-\frac{q}{q -1}}.
\end{equation} 
where the chemical potential does not appear explicitly.

It thus requires special attention to determine the chemical potential using the above Tsallis distribution when fitting 
experimental data.

It is to be noted that most previous analyses have confused the two equations~(\ref{YieldNonZeroMu}) and (\ref{YieldIndexZero}) and 
reached
conclusions that are incorrect,  namely  that at LHC energies, different hadrons,
$\pi, K, p, ...$ cannot be described by the same values of $T$ and $V$. As we will show this is based on using $T_0$ and $V_0$
and not $T$ and $V$. Many authors have followed this conclusion because at LHC energies equal numbers of
particles and antiparticles are being produced and, furthermore, at chemical equilibrium, one has indeed $\mu = 0$
for all quantum numbers. However the equality 
of particle and antiparticle yields, at thermal freeze-out,  only implies that e.g. $\pi^+$ and $\pi^-$ have the same chemical 
potential but they are  not necessarily zero.

It is the purpose of the present paper to resolve this issue.
The procedure we choose is the following:
\begin{enumerate}
\item Use Eq.~(\ref{YieldIndexZero}) to fit the transverse momentum distributions. This determines the three parameters $T_0$, $q$ and $V_0$.
\item Fix the parameter $q$ thus obtained.
\item Perform a new fit to the transverse momentum distributions using Eq.~(\ref{Yield}) keeping $q$ as determined in the previous step.
This determines the parameters $T$ and $V$ and the chemical potential $\mu$.
\item Check the consistency with Eqs.~(\ref{T0}) and (\ref{V0}).
\end{enumerate}
Each step in the fitting procedure thus  involves only three parameters to describe the transverse momentum distributions.

We emphasize that the chemical potentials at kinetic freeze-out (described here with a Tsallis distribution), are not related to those
 at chemical freeze-out. 
At chemical freeze-out, where thermal and chemical equilibrium have been well established 
the chemical potentials are zero.
At kinetic freeze-out however, there is no chemical equilibrium and the observed particle-antiparticle symmetry only
implies that the chemical potentials for particles must be equal to those for antiparticles. 
However, due to the absence of chemical  equilibrium they do not have to be zero.
 The only constraint is that they should be equal for particles and antiparticles.

As mentioned above,  the advantage of using the above distribution is that they follow consistent set of thermodynamic relations 

\begin{eqnarray}\label{Thermodynamic_relations}
d\epsilon  & = & T\,ds  + \mu\, dn,\\
dP              & = & s\,dT  + nd\mu,
\end{eqnarray}
namely, with the distribution
\begin{equation}
 f(E,q,T,\mu) \equiv \left[1+(q-1)\ \frac{E - \mu}{T}\right]^{-\frac{1}{q-1}},
\label{tsallis}
\end{equation}
and 
\begin{eqnarray}
n &\equiv&\,g \int \frac{d^3p}{(2\pi)^3} f^q \label{Number},\\
\epsilon \,&\equiv&\,g \int \frac{d^3p}{(2\pi)^3}E f^q \label{epsilon},\\
s &=& - g\int\frac{d^3p}{(2\pi)^3}\left[f^{q}{\rm ln}_{q}f - f\right],\label{entropy}\\
P\,&\equiv&\,g \int \frac{d^3p}{(2\pi)^3}\frac{p^2}{3~E}\ f^q \label{pressure}.
\end{eqnarray}
where $\epsilon$ is the energy density, $T$ is the temperature, $s$ is the entropy density, $P$ is the pressure, $\mu$ is the chemical potential and $n$ is the particle density.  It was shown (see~\cite{worku1, worku2,cleymans2016large} for more details) that   $n, \epsilon, s$ and  $P$ given by the following relations:

It is thus clear that the parameter $T$ can now  now considered as a temperature in the thermodynamic sense:
\begin{equation}
  T = \left.\frac{\partial E}{\partial S}\right|_{V,N}, 
\end{equation}
where the entropy $S$ is the Tsallis entropy. 
 
It is the purpose of the present paper to determine the chemical potential using the above Tsallis distribution when fitting experimental data.
%
\section{Fits without chemical potentials using Eq.~(\ref{YieldIndexZero}).}

The Tsallis distribution has been widely used in the analysis of transverse momentum
 spectra~\cite{bialas,grigoryan,wong,ristea,biro,parvan1,parvan2,tripathy,zheng1,marques,azmi,sorin,sena,Azmi:2014dwa,Biro:2004qg,Khuntia:2017ite,zheng,de}. 
Most of these  put the chemical potential equal to zero and therefore use Eq.~(\ref{YieldIndexZero}). We can therefore take over their results 
which we reinterpret as being for $T_0$, $V_0$ and $q$. The results  obtained are reproduced in Table~1. No new fits are made
and we base our results on those reported in a recent analysis~\cite{bhattacharyya2018precise}.
It is to be noted that e.g. in $p-p$ collisions at $\sqrt{s} = 7$ TeV the radius varies between 3 and 5.7 fm while 
the temperature $T_0$ varies between 66 and 101 MeV.
 
\section{Tsallis Fits with Chemical Potential Using Eq.~(\ref{YieldNonZeroMu})}

\begin{table}[!ht]
\centering
	\label{tab:results_fixedmu}
\centering
\begin{tabular}{|c|c|c|c|c|c|}\hline
$\sqrt{s}$ (TeV)&Particle     & $R_0$ (fm)             & $q$                   & $T_0$ (GeV)      & $\chi^2$ / NDF          \\ 
\hline
0.9~\cite{ALICE2} &$\pi^+$ & 4.835 $\pm$ 0.136  & 1.148 $\pm$ 0.005 & 0.070 $\pm$ 0.002 & 22.73 / 30  \\
&$\pi^-$   & 4.741 $\pm$ 0.131     & 1.145 $\pm$ 0.005     & 0.072 $\pm$ 0.002    & 15.83 / 30  \\
&$K^+$     & 4.523 $\pm$ 1.302    & 1.175 $\pm$ 0.017     & 0.057 $\pm$ 0.013    & 13.02 / 24    \\
&$K^-$     & 3.957 $\pm$ 0.962    & 1.161 $\pm$ 0.016     & 0.064 $\pm$ 0.013    & 6.214/ 24   \\
&$p$       & 42.72 $\pm$ 19.8	 &  1.158 $\pm$ 0.006   &  0.020 $\pm$ 0.004  &  14.29/21 \\
&$\bar{p}$ & 7.445 $\pm$ 3.945   & 1.132 $\pm$ 0.014    & 0.052 $\pm$ 0.016     & 13.82/ 21 \\
\hline
2.76~\cite{ALICE_2760} &$\pi^+ + \pi^-$ & 4.804 $\pm$ 0.100 & 1.149 $\pm$ 0.002 & 0.077 $\pm$ 0.001    & 20.64 / 60  \\
&$K^+ + K^-$     & 2.51 $\pm$ 0.128  & 1.144 $\pm$ 0.002     & 0.096 $\pm$ 0.004  & 2.459 /55    \\
&$p + \bar{p}$ & 4.009 $\pm$ 0.623   & 1.121 $\pm$ 0.005    & 0.086 $\pm$ 0.008     & 3.509 / 46 \\
\hline
5.02~\cite{ALICE_5020} &$\pi^+ + \pi^-$ & 5.025 $\pm$ 0.111 & 1.155 $\pm$ 0.002 & 0.076 $\pm$ 0.002    & 20.13 / 55  \\
&$K^+ + K^-$     & 2.437 $\pm$ 0.168  & 1.15 $\pm$ 0.005     & 0.099 $\pm$ 0.006  & 1.516 /48    \\
&$p + \bar{p}$ & 3.601 $\pm$ 0.546   & 1.126 $\pm$ 0.005    & 0.091 $\pm$ 0.009     & 2.558 / 46 \\
\hline
7.0~\cite{ALICE_5020,ALICE_7000} &$\pi^+ + \pi^-$  & 5.664 $\pm$ 0.167 & 1.179 $\pm$ 0.003 & 0.066 $\pm$ 0.002   & 14.14 /38  \\
&$K^+ + K^-$     & 2.511 $\pm$ 0.145  & 1.158 $\pm$ 0.005 & 0.097 $\pm$ 0.005    & 3.114 /45    \\
&$p + \bar{p}$ & 3.074 $\pm$ 0.405   & 1.124 $\pm$ 0.005  & 0.101 $\pm$ 0.008   & 6.031 / 43 \\
\bottomrule
\hline
\end{tabular}
\caption{Fit results at $\sqrt{s}$ = 900 GeV ~\cite{ALICE2}, $\sqrt{s}$ = 2.76 TeV ~\cite{ALICE_2760}, 5.02 TeV ~\cite{ALICE_5020} and 7 TeV~\cite{ALICE_5020,ALICE_7000}, using data from the ALICE collaboration, with $\mu$ fixed to zero using Eq.~(\ref{YieldIndexZero}).}
\end{table}

\begin{table}[!ht]
	\centering
	\label{tab:results_fixedq}
	\centering
	\begin{tabular}{|c|c|c|c|c|c|}\hline
		$\sqrt{s}$ (TeV)&Particle     & $R$ (fm)             & $\mu$ (GeV)                  & $T$ (GeV)      & $\chi^2$ / NDF          \\ \hline
		0.9~\cite{ALICE2} &$\pi^+$ & 4.228 $\pm$ 0.276  & 0.159 $\pm$ 0.014 & 0.094 $\pm$ 0.002 & 23.018 / 30  \\
		&$\pi^-$   & 4.34 $\pm$ 0.256  & 0.151 $\pm$ 0.015   & 0.094 $\pm$ 0.002  & 15.776 / 30  \\
		&$K^+$     & 4.313 $\pm$ 10.5  &  0.112$\pm$ 0.476   & 0.077 $\pm$ 0.077  & 13.019 /23    \\
		&$K^-$     & 4.367 $\pm$ 0.362 & 0.095 $\pm$ 0.027   & 0.080 $\pm$ 0.004 & 6.214 / 23   \\
		&$p$      & 3.922  $\pm$ 4.3   &   0.314 $\pm$ 0.216   & 0.067 $\pm$ 0.030 & 14.519/20  \\
		&$\bar{p}$ & 3.903 $\pm$ 0.296 &  0.218 $\pm$ 0.032  & 0.081  $\pm$ 0.003  & 13.830 /20   \\
		\hline
		2.76~\cite{ALICE_2760} &$\pi^+ + \pi^-$   & 4.216 $\pm$ 2.521   & -0.088$\pm$ 0.100     & 0.064  $\pm$ 0.015   &  20.48/60   \\
		&$K^+ + K^-$    & 4.415 $\pm$ 3.783 & -0.203 $\pm$ 0.130  & 0.058  $\pm$ 0.019    & 7.62    /55   \\
		&$p + \bar{p}$ & 4.051 $\pm$ 5.033   & -0.126 $\pm$ 0.234    & 0.070 $\pm$ 0.028     & 3.518/46 \\
		\hline
		5.02~\cite{ALICE_5020} &$\pi^+ + \pi^-$   & 4.394 $\pm$ 2.756 & 0.027$\pm$ 0.130 & 0.080 $\pm$ 0.020   & 20.14 /55   \\
		&$K^+ + K^-$    & 4.617 $\pm$ 5.828  & -0.148 $\pm$ 0.255  & 0.078 $\pm$ 0.038    & 1.522 /48   \\
		&$p + \bar{p}$ & 4.415 $\pm$ 5.191    & -0.046$\pm$ 0.267  & 0.085 $\pm$ 0.033   &  2.561/46   \\
		\hline
	7.0~\cite{ALICE_5020,ALICE_7000}  &$\pi^+ + \pi^-$   & 4.178 $\pm$ 0.287   & 0.192$\pm$ 0.018    & 0.100  $\pm$ 0.003   &  14.15/38   \\
		&$K^+ + K^-$    & 4.205 $\pm$ 0.017 & 0.023 $\pm$ 0.005  & 0.102  $\pm$ 0.001    & 3.128    /55   \\
		&$p + \bar{p}$ & 4.43 $\pm$ 0.298   & 0.070 $\pm$ 0.022 & 0.110 $\pm$ 0.003     & 6.031/43 \\
		\bottomrule
\hline
	\end{tabular}
\caption{Fit results at $\sqrt{s}$ = 900 GeV ~\cite{ALICE2}, $\sqrt{s}$ = 2.76 TeV ~\cite{ALICE_2760}, 
5.02 TeV~\cite{ALICE_5020} and 7 TeV~\cite{ALICE_5020,ALICE_7000}, using data from the 
ALICE collaboration with $q$ from Table~1 following Eq.~(\ref{YieldNonZeroMu}).}
	
\end{table}

The fits to the transverse momentum distributions were then repeated using Eq.~(\ref{YieldNonZeroMu}) but this time keeping the parameter $q$ fixed 
to the value 
determined in the previous section and listed in Table~1. The results are listed 
 in Table~2, where we present the fit results for non-zero chemical potential
 for $p-p$ collisions at four different beam energies by the 
ALICE collaboration.  

It is to be noted that the entry for the proton at 900 GeV has a very large uncertainty; for this reason we also considered
the results obtained by the CMS collaboration~\cite{CMS_7000} at the LHC.
The results  are shown in Tables 3 and 4, in this case the proton can be determined more accurately.

\begin{table}[!ht]
\centering
	\label{tab:results_CMS_fixedmu}
\centering
\begin{tabular}{|c|c|c|c|c|c|}\toprule
$\sqrt{s}$ (TeV)&Particle     & $R_0$ (fm)             & $q$                   & $T_0$ (GeV)      & $\chi^2$ / NDF          \\ \midrule
0.9~\cite{CMS_7000} &$\pi^+$ & 4.312 $\pm$ 0.123  & 1.164 $\pm$ 0.005 & 0.077 $\pm$ 0.002 & 4.044 / 19  \\
&$\pi^-$   & 5.449 $\pm$ 0.158     & 1.167 $\pm$ 0.005     & 0.066 $\pm$ 0.002    & 10.3 / 19  \\
&$K^+$     & 3.297 $\pm$ 0.984    & 1.158 $\pm$ 0.036     & 0.078 $\pm$ 0.022    & 2.123 / 14    \\
&$K^-$     & 4.053 $\pm$ 1.918    & 1.182 $\pm$ 0.046     & 0.064 $\pm$ 0.027    & 1.236/ 14   \\
&$p$       & 6.118 $\pm$ 3.725	 &  1.139 $\pm$ 0.020   &  0.058 $\pm$ 0.022  &  9.596/24 \\
&$\bar{p}$ & 8.619 $\pm$ 0.211   & 1.147 $\pm$ 0.001    & 0.047 $\pm$ 0.007     & 21.3/ 24 \\
\hline
2.76~\cite{CMS_7000} &$\pi^+$ & 6.195 $\pm$ 0.201  		& 1.189 $\pm$ 0.005 & 0.061 $\pm$ 0.002 & 5.711 / 19  \\
&$\pi^-$   & 5.971 $\pm$ 0.186     & 1.184 $\pm$ 0.005     & 0.063 $\pm$ 0.002    & 7.077 / 19  \\
&$K^+$     & 2.997 $\pm$ 0.826    & 1.162 $\pm$ 0.040     & 0.087 $\pm$ 0.024    & 2.447 / 14    \\
&$K^-$     & 2.683 $\pm$ 0.685    & 1.147 $\pm$ 0.041     & 0.096 $\pm$ 0.024    & 7.407 / 14   \\
&$p$       & 7.192 $\pm$ 0.170	  & 1.166 $\pm$ 0.002   &  0.049 $\pm$ 0.001     &  27.43 /24 \\
&$\bar{p}$ & 3.028 $\pm$ 1.167    & 1.129 $\pm$ 0.025    & 0.087 $\pm$ 0.025     & 28.41 / 24 \\
\hline
7.0~\cite{CMS_7000} &$\pi^+$ & 6.762 $\pm$ 0.246  & 1.203 $\pm$ 0.006 & 0.059 $\pm$ 0.002 & 14.29 / 19  \\
&$\pi^-$   & 6.614 $\pm$ 0.233     & 1.202 $\pm$ 0.006     & 0.060 $\pm$ 0.002    & 11.36 / 19  \\
&$K^+$     & 2.642 $\pm$ 0.588    & 1.152 $\pm$ 0.041     & 0.102 $\pm$ 0.024    & 2.074 / 14    \\
&$K^-$     & 3.221 $\pm$ 1.422    & 1.186 $\pm$ 0.063     & 0.083 $\pm$ 0.036    & 4.38 / 14   \\
&$p$       & 6.076 $\pm$ 0.145	 &  1.184 $\pm$ 0.002   &  0.052 $\pm$ 0.001  &  12.22 /24 \\
&$\bar{p}$ & 7.394 $\pm$ 0.178   & 1.190 $\pm$ 0.002    & 0.045 $\pm$ 0.001     & 15.47 / 24 \\
\hline
13.0~\cite{CMS13} &$\pi^+$ & 6.719 $\pm$ 0.305		  & 1.215 $\pm$ 0.008	 & 0.057 $\pm$ 0.003 & 3.546 / 19  \\
&$\pi^-$   & 5.785 $\pm$ 0.222     & 1.191 $\pm$ 0.008    & 0.067 $\pm$ 0.003    & 12.72 / 19  \\
&$K^+$     & 2.477 $\pm$ 0.071    & 1.142 $\pm$ 0.071     & 0.106 $\pm$ 0.041    & 1.828 / 14    \\
&$K^-$     & 2.566 $\pm$ 1.667    & 1.155 $\pm$ 0.114     & 0.100 $\pm$ 0.065    & 1.323 / 14   \\
&$p$       & 17.34 $\pm$ 7.603    &  1.206 $\pm$ 0.007     & 0.025 $\pm$ 0.007    & 8.921 /24 \\
&$\bar{p}$ & 6.516 $\pm$ 0.239    & 1.189 $\pm$ 0.003      & 0.048 $\pm$ 0.001     & 8.383 / 24 \\
\bottomrule
\hline
\end{tabular}
\caption{Fit results using Eq.~(\ref{YieldIndexZero}) at $\sqrt{s} $ = 900 GeV, $\sqrt{s}$ = 2.76 TeV, 5.02 TeV and 7 TeV, using data from the CMS collaboration.}
\end{table}

\begin{table}[!ht]
	\centering
	\label{tab:results_CMS_fixedq}
	\centering
	\begin{tabular}{|c|c|c|c|c|c|}\toprule
		$\sqrt{s}$ (TeV)&Particle     & $R$ (fm)             & $\mu$ (GeV)                  & $T$ (GeV)      & $\chi^2$ / NDF          \\ \midrule
		0.9~\cite{CMS_7000} &$\pi^+$ & 4.007 $\pm$ 0.104  & 0.137 $\pm$ 0.006 & 0.089 $\pm$ 0.001 & 4.056 / 19  \\
		&$\pi^-$   & 4.005 $\pm$ 0.014  & 0.191 $\pm$ 0.002   & 0.099 $\pm$ 0.001  & 10.74 / 19  \\
		&$K^+$     & 4.033 $\pm$ 0.138  &  0.089  $\pm$ 0.010   & 0.092 $\pm$ 0.001  & 2.123 /14    \\
		&$K^-$     & 4.392 $\pm$ 0.015  & 0.097 $\pm$ 0.009   & 0.081 $\pm$ 0.001 & 1.236 / 14   \\
		&$p$      & 4.184  $\pm$ 0.126  &   0.184 $\pm$ 0.009   & 0.084 $\pm$ 0.001 & 9.596/24  \\
		&$\bar{p}$ & 4.112 $\pm$ 0.124 &  0.219 $\pm$ 0.009  & 0.079  $\pm$ 0.001  & 21.3 /24   \\
		\hline
		2.76~\cite{CMS_7000} &$\pi^+$ & 3.999 $\pm$ 0.103       & 0.211 $\pm$ 0.007 & 0.101 $\pm$ 0.001 & 5.712 / 19  \\
		&$\pi^-$   & 4.003 $\pm$ 0.102  & 0.208 $\pm$ 0.007   & 0.102 $\pm$ 0.001  & 7.077 / 19  \\
		&$K^+$     & 3.98 $\pm$ 0.136  &  0.079  $\pm$ 0.011   & 0.100 $\pm$ 0.001  & 2.447 /14    \\
		&$K^-$     & 4.048 $\pm$ 0.014  & 0.053 $\pm$ 0.011   & 0.104 $\pm$ 0.002 & 7.407 / 14   \\
		&$p$      & 4.251  $\pm$ 0.025  &   0.186 $\pm$ 0.008   & 0.080 $\pm$ 0.001 & 27.43 /24  \\
		&$\bar{p}$ & 4.198 $\pm$ 0.124 &  0.073 $\pm$ 0.011  & 0.099  $\pm$ 0.001  & 28.41 /24   \\
			\hline
		7.0~\cite{CMS_7000} &$\pi^+$ & 4.032 $\pm$ 0.104  & 0.204 $\pm$ 0.008     & 0.118 $\pm$ 0.001 & 235.7 / 19  \\
		&$\pi^-$   & 4.020 $\pm$ 0.102  & 0.202 $\pm$ 0.008   & 0.119 $\pm$ 0.001  & 240.4 / 19  \\
		&$K^+$     & 4.144 $\pm$ 0.144  &  0.045  $\pm$ 0.012   & 0.109 $\pm$ 0.002  & 2.074 /14    \\
		&$K^-$     & 3.987 $\pm$ 0.140  & 0.092 $\pm$ 0.011   & 0.100 $\pm$ 0.002 & 4.38 / 14   \\
		&$p$      & 4.324  $\pm$ 0.025  &  0.160 $\pm$ 0.009   & 0.081 $\pm$ 0.001 & 12.22 /24  \\
		&$\bar{p}$ & 4.418 $\pm$ 0.125 &  0.171 $\pm$ 0.009  & 0.078  $\pm$ 0.001  & 15.47 /24   \\
			\hline
		13.0~\cite{CMS13} &$\pi^+$ & 3.948 $\pm$ 0.154  & 0.207 $\pm$ 0.012 & 0.116 $\pm$ 0.001 & 138 / 19  \\
		&$\pi^-$   & 4.023 $\pm$ 0.015  & 0.211 $\pm$ 0.002   & 0.107 $\pm$ 0.001  & 12.72 / 19  \\
		&$K^+$     & 3.990 $\pm$ 0.264  &  0.039  $\pm$ 0.023   & 0.111 $\pm$ 0.003  & 1.828 /14    \\
		&$K^-$     & 3.997 $\pm$ 0.267  & 0.045 $\pm$ 0.022   & 0.107 $\pm$ 0.004 & 1.323 / 14   \\
		&$p$      & 4.479  $\pm$ 0.194  & -0.062 $\pm$ 0.020   & 0.112 $\pm$ 0.002 & 15.95 /24  \\
		&$\bar{p}$ & 4.410 $\pm$ 0.189 &  0.155 $\pm$ 0.015  & 0.078  $\pm$ 0.001  & 80383 /24   \\
		\bottomrule
		\hline
	\end{tabular}
\caption{Fit results  using Eq.~(\ref{YieldNonZeroMu}) at $\sqrt{s} $ = 900 GeV, $\sqrt{s}$ = 2.76 TeV, 5.02 TeV and 7 TeV, using data 
from the CMS collaboration with $q$ from Table~\ref{tab:results_fixedmu}.}
	
\end{table}

The conclusions remain unchanged: the results including the chemical potential lead to more consistent values
for the temperature $T$ and the radius $R$.

\begin{figure}[!ht]
	\centering
	\includegraphics[height = 12cm, width = 12cm]{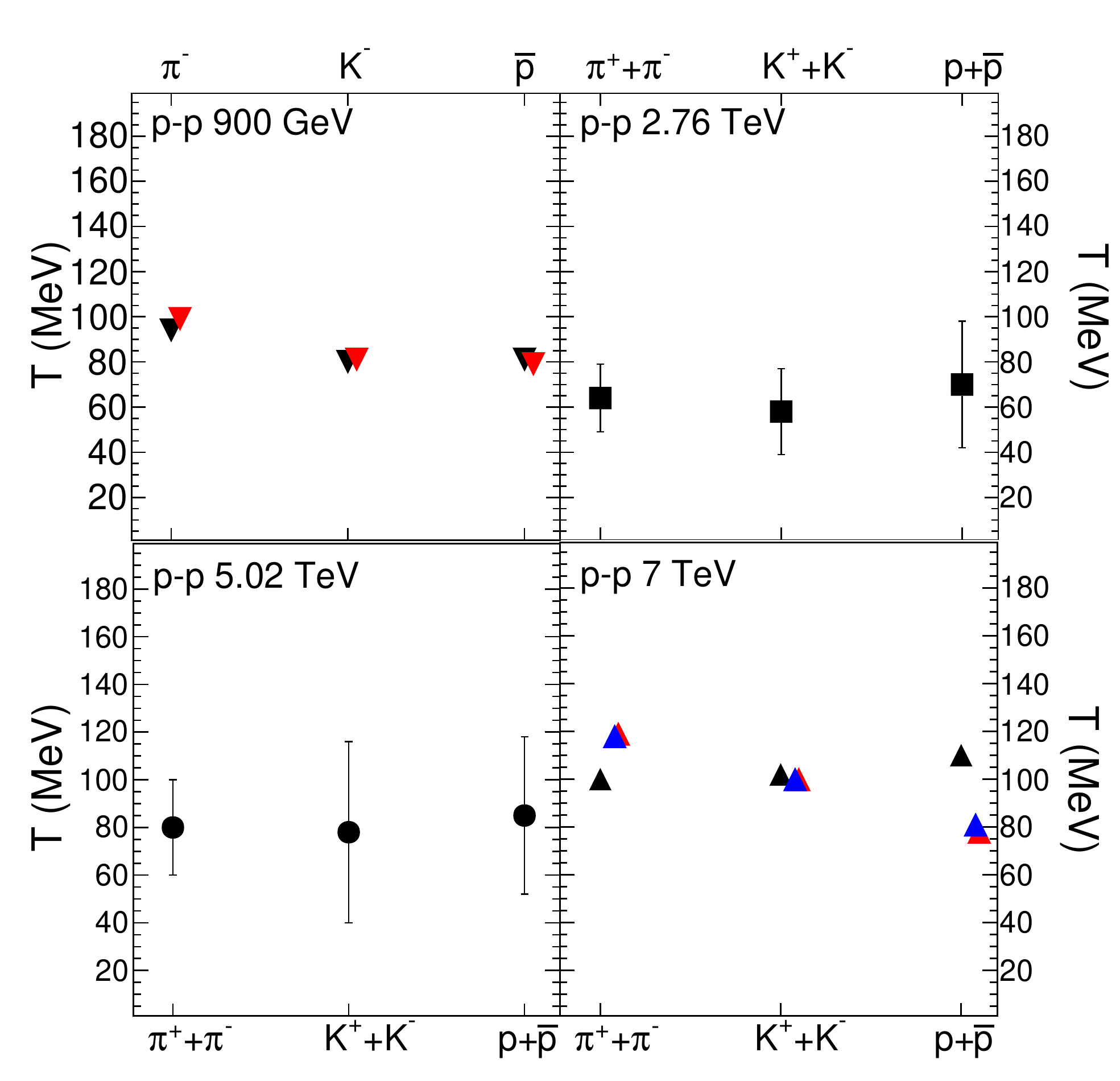} %
	\caption{\label{T_allenergies}Values of $T$  
obtained by fitting data in $p-p$ collisions using  Eq.~(\ref{YieldNonZeroMu}) . 
The upper-left plot was obtained using data at 900 GeV where the black triangles were obtained using data from the ALICE 
collaboration~\cite{ALICE2} while the red triangles use data from the CMS collaboration~\cite{CMS_7000}. The upper-right
plot uses data at 2.76 TeV~\cite{ALICE_2760}, the lower-right plot is for 5.02 TeV~\cite{ALICE_5020} while the bottom-right plot
is from $p-p$ data at 7 TeV~\cite{ALICE_7000}. In the last case we also show the results obtained from the CMS data~\cite{CMS_7000}, red is for negative
hadrons, blue for positive ones.  
} 
\end{figure}

\begin{figure}[!ht]
	\centering
	\includegraphics[height = 12cm, width = 12cm]{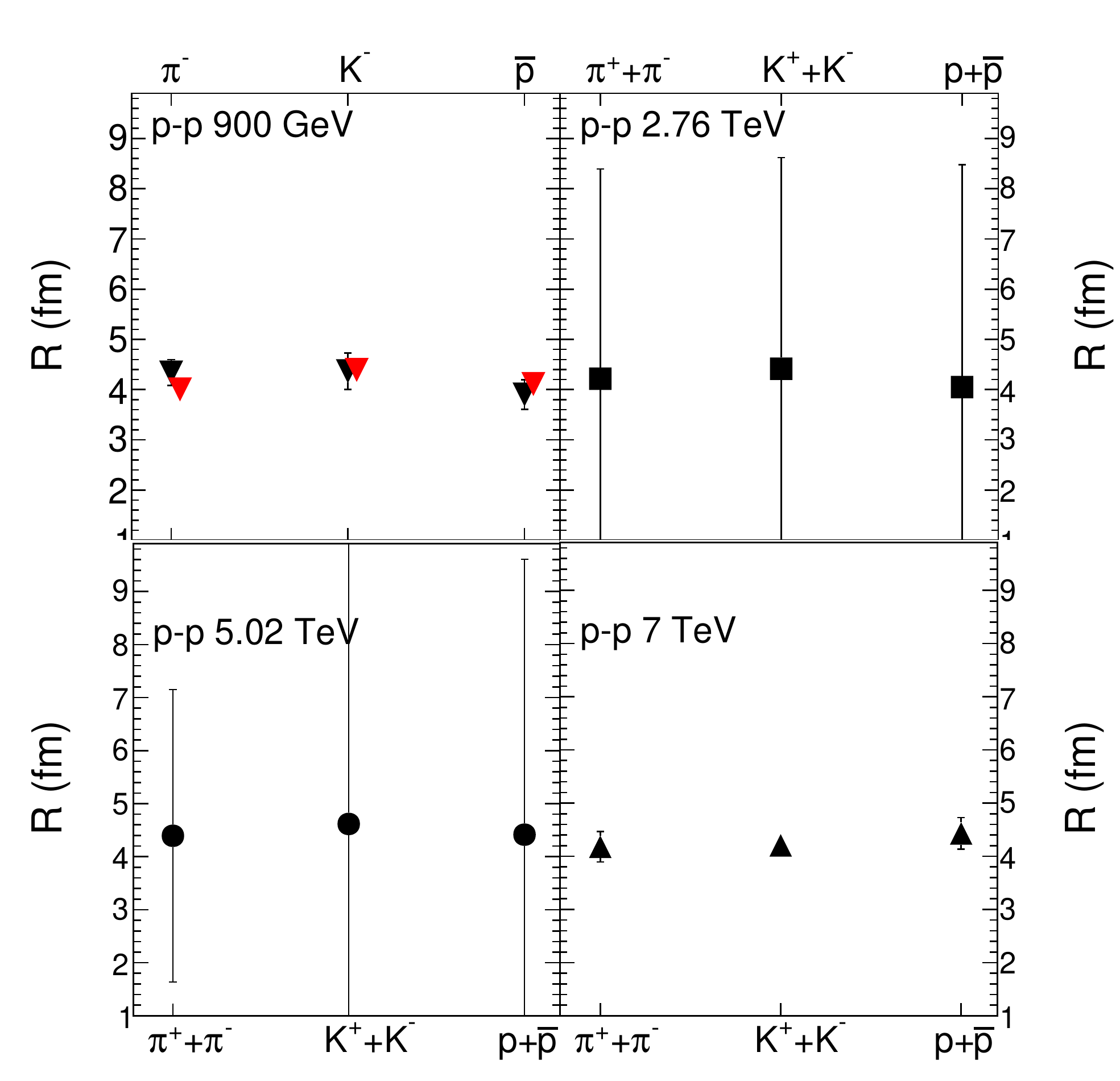} %
	\caption{\label{R_allenergies} Radius as a function of particle type for four different beam energies 
in $p-p$ collisions obtained using Eq.~(\ref{YieldNonZeroMu}) at mid-rapidity.
The upper-left plot was obtained using data at 900 GeV where the black triangles were obtained using data from the ALICE 
collaboration~\cite{ALICE2} while the red triangles use data from the CMS collaboration~\cite{CMS_7000}. the upper-right
plot uses data at 2.76 TeV~\cite{ALICE_2760}, the lower-right plot is for 5.02 TeV~\cite{ALICE_5020} while the bottom-right plot
is from $p-p$ data at 7 TeV~\cite{ALICE_7000}.  
}
\end{figure}

Focusing again on  $p-p$ collisions at $\sqrt{s} = 7$ TeV the radius varies between 3 and 5.7 fm, the radius now varies between 4.18 and 4.4 fm
instead of between 3 and 5.7 fm. The radii are now consistent with each other within error bars. The temperature $T$ now 
varies between 100 and 110 MeV while previously the variation was between  between 66 and 101 MeV.  In general all values obtained are
much more consistent with each other. This completely changes the overall picture obtained for the parameters obtained using the Tsallis distribution.
The results for the temperature $T$ are shown in Fig.~\ref{T_allenergies}. The uncertainties obtained at beam energies of 2.76 TeV and 5.02 TeV 
are large, it is to be hoped that these will be reduced in future analyses of the data.
The results for the radii are shown in Fig.~\ref{R_allenergies} where a similar picture emerges, namely radii at each beam energy  are very
similar for all the particle types considered.  This lends support to the picture that the final state at kinetic freeze-out is in thermal
equilibrium albeit in the Tsallis sense of equilibrium.

\begin{figure}[!ht]
	\centering
	\includegraphics[height = 12cm, width = 12cm]{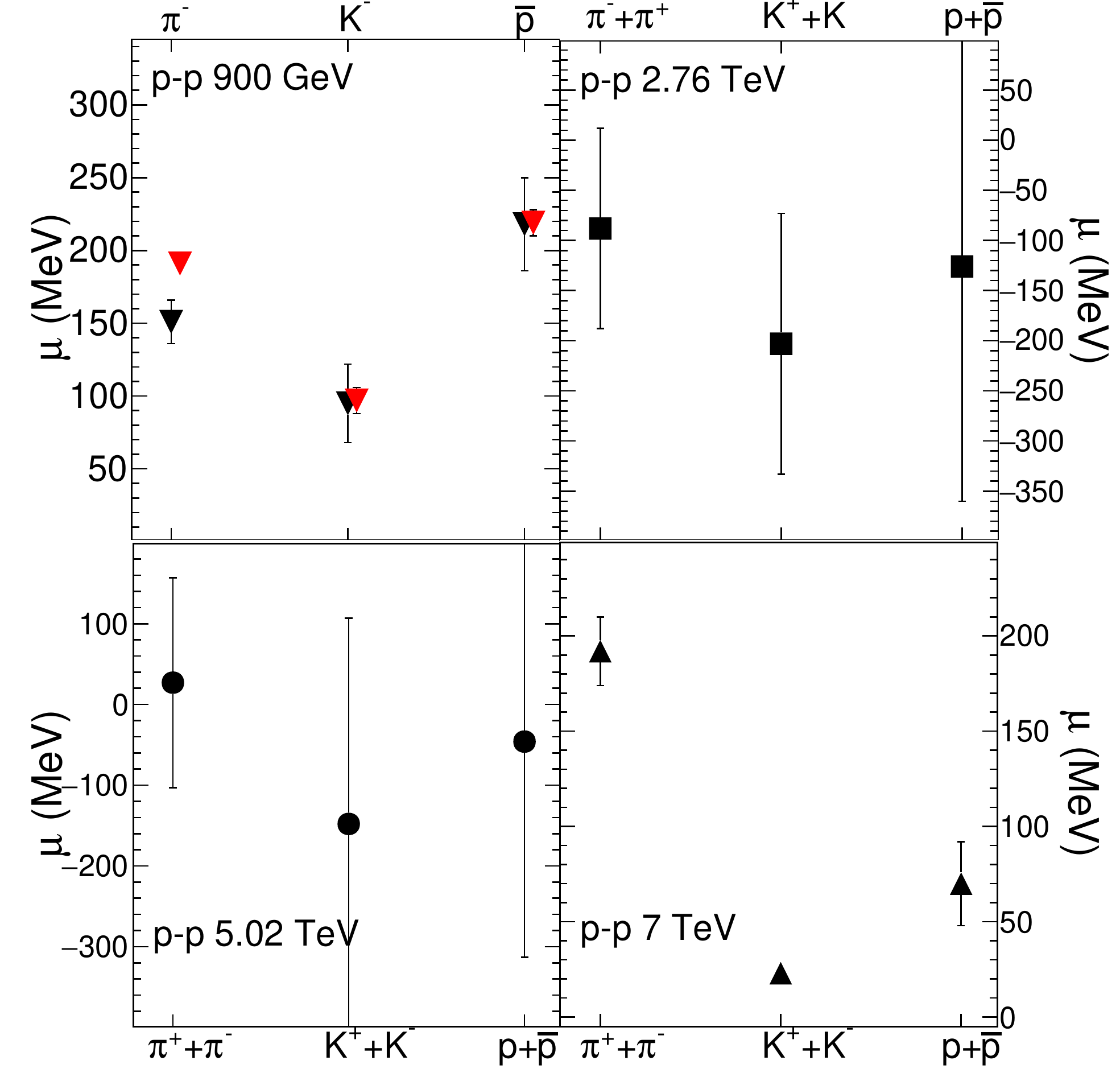} %
	\caption{\label{mu_allenergies}Chemical potential as a function of particle type. 
The upper-left plot was obtained using data at 900 GeV where the black triangles were obtained using data from the ALICE 
collaboration~\cite{ALICE2} while the red triangles use data from the CMS collaboration~\cite{CMS_7000}. the upper-right
plot uses data at 2.76 TeV~\cite{ALICE_2760}, the lower-right plot is for 5.02 TeV~\cite{ALICE_5020} while the bottom-right plot
is from $p-p$ data at 7 TeV~\cite{ALICE_7000}.  
}
\end{figure}

Finally we show the chemical potentials at kinetic freeze-out as listed in Table~2 also in Fig.~\ref{mu_allenergies}.
We find that the chemical potentials are not always compatible with zero and can even be quite large. For example, for $p-p$ collisions at
7 TeV the chemical potential for pions is around 200 MeV while for kaons it is close to 20 MeV  and for protons  it is around 70 MeV. They are thus
quite large, clearly non-zero and also very different from each other. This does not come as a surprise as no chemical equilibrium is expected 
to exist at kinetic freeze-out in $p-p$ collisions. We find the near equality of the values for $T$ and $R$ very interesting.

We have also checked the 
 consistency of our results by  comparing explicitly the values obtained for
 $T_0$ obtained by making use of Eq.~(\ref{YieldIndexZero}) to those obtained by fitting the transverse momentum spectra 
using Eq.~({\ref{YieldNonZeroMu}) combined with Eq.~(\ref{T0}). The results are indeed consistent with each other within
the errors of the analysis.
This further confirms an assertion by~\cite{sorin} that there is a redundancy in the parameters appearing in Eq.~($\ref{YieldNonZeroMu}$). 

\section{Summary}
A comparison of $T$ and $T_0$ values for all the energies considered, for both the ALICE and CMS Collaborations  are in agreement. This result confirms that the  variables $T,V,q,$ and $\mu$ in the Tsallis distribution function Eq.~$(\ref{Yield})$ have a redundancy for $\mu \neq 0$~\cite{sorin}.\\
The conclusions are as follows:
\begin{itemize}
\item The Tsallis distribution gives a good description of the transverse momentum spectra in $p-p$ collisions. These 
correspond to the thermal freeze-out stage in the collisions at LHC energies.
\item There is a reasonable amount of thermal equilibrium even at the thermal freeze-out stage, albeit in the Tsalllis
thermodynamics sense. The resulting temperatures $T$ are the same for all particle types considered. This is in contrast to the results 
obtained in~\cite{bhattacharyya2018precise}, the difference can be traced back to the incorrect use of chemical potentials in the latter analysis.
\item The values obtained for the volume (or the radius shown in Fig.~\ref{R_allenergies}) are consistent with being equal 
for all particle types considered here. Admittedly the error bars are large at beam energies of 2.76 and 5.02 TeV for $p-p$ 
collisions.
\item There is no chemical equilibrium at thermal freeze-out; the values obtained for $\mu$ for different particle types
vary considerably. 
\end{itemize}
In summary, in this work we have correctly taken into account the chemical potential in the Tsallis distribution analysis
of $p-p$ collisions at four different LHC energies, this is summarized in the distinction that has to be made
between $T_0$ as defined in Eq.~(\ref{YieldIndexZero}) and $T$ as defined in the starting Eq.~(\ref{YieldNonZeroMu}). This leads to a much
more satisfactory analysis of the various parameters and confirms the usefulness of the Tsallis distribution in
high-energy collisions.
\section*{References}

\bibliographystyle{elsarticle-num}
\bibliography{cp}
\end{document}